\documentclass[aps,preprint]{revtex4}%
\usepackage{revsymb}
\usepackage{amsmath}
\usepackage{graphicx}%
\usepackage{amsfonts}%
\usepackage{amssymb}
\newtheorem{theorem}{Theorem}
\newtheorem{acknowledgement}[theorem]{Acknowledgement}

\begin{document}

\preprint{Publica\c{c}\~{a}o IFUSP 1607/2005}
\title{{\Large Symmetries of dynamically equivalent theories}}
\author{D.M. Gitman and I.V. Tyutin}
\affiliation{Institute of Physics, University of S\~{a}o Paulo, Brazil; e-mail: gitman@fma.if.usp.br}
\affiliation{Lebedev Physics Institute, Moscow, Russia; e-mail: tyutin@lpi.ru}

\begin{abstract}
A natural and very important development of constrained system theory is a
detail study of the relation between the constraint structure in the
Hamiltonian formulation with specific features of the theory in the Lagrangian
formulation, especially the relation between the constraint structure with the
symmetries of the Lagrangian action. An important preliminary step in this
direction is a strict demonstration, and this is the aim of the present
article, that the symmetry structures of the Hamiltonian action and of the
Lagrangian action are the same. This proved, it is sufficient to consider the
symmetry structure of the Hamiltonian action. The latter problem is, in some
sense, simpler because the Hamiltonian action is a first-order action. At the
same time, the study of the symmetry of the Hamiltonian action naturally
involves Hamiltonian constraints as basic objects. One can see that the
Lagrangian and Hamiltonian actions are dynamically equivalent. This is why, in
the present article, we consider from the very beginning a more general
problem: how the symmetry structures of dynamically equivalent actions are
related. First, we present some necessary notions and relations concerning
infinitesimal symmetries in general, as well as a strict definition of
dynamically equivalent actions. Finally, we demonstrate that there exists an
isomorphism between classes of equivalent symmetries of dynamically equivalent actions.

\end{abstract}

\maketitle
\tableofcontents

\section{Introduction}

The most of contemporary particle-physics theories are formulated as gauge
theories. It is well known that within the Hamiltonian formulation gauge
theories are theories with constraints. This is the main reason for a long and
intensive study of the formal theory of constrained systems, see \cite{1}. It
still attracts considerable attention of researchers. From the very beginning,
it became clear that the presence of first-class constraints among the
complete set of constraints in the Hamiltonian formulation is a direct
indication that the theory is a gauge one, i.e., its Lagrangian action is
invariant under gauge transformations. A next natural, and very important,
step would be a detail study of the relation between the constraint structure
and constraint dynamics in the Hamiltonian formulation with specific features
of the theory in the Lagrangian formulation, especially the relation between
the constraint structure with the gauge transformation structure of the
Lagrangian action. An important problem to be solved in this direction would
be a strict demonstration, and this is the aim of the present article, that
the symmetry structures of the Hamiltonian action and of the Lagrangian action
are the same. This proved, it is sufficient to consider the symmetry structure
of the Hamiltonian action. The latter problem is, in some sense, simpler
because the Hamiltonian action is a first-order action. At the same time, the
study of the symmetry of the Hamiltonian action naturally involves Hamiltonian
constraints as basic objects, see \cite{2,3}. It follows from the results of
the article \cite{4} that the Lagrangian and Hamiltonian actions are
dynamically equivalent. This is why in the present article we consider from
the very beginning a more general problem: how the symmetry structures of
dynamically equivalent actions are related. The article is organized as
follows: In sec. 2, we present some necessary notions and relations concerning
infinitesimal symmetries in general. A strict definition of dynamically
equivalent actions is given in sec. 3. Finally, in sec. 4, we demonstrate that
there exists an isomorphism between classes of equivalent symmetries of
dynamically equivalent actions.

\section{Symmetries}

\subsection{Basic notation and relations}

We consider finite-dimensional systems which are described by the generalized
coordinates\emph{ }$q\equiv\{q^{a};\;a=1,2,...,n\}$. The space of the
variables $q^{a\left[  l\right]  },$%
\begin{equation}
q^{a\left[  l\right]  }=\left(  d_{t}\right)  ^{l}q^{a}\,,\;\;l=0,1,...,N_{a}%
,\;\;\left(  q^{a\left[  0\right]  }=q^{a}\right)  \,,\;d_{t}=\frac{d}{dt}\,,
\label{1.1}%
\end{equation}
\ considered as independent variables, with finite $N_{a}$ , or with some
infinite $N_{a}\,,$ is called the jet space. The majority of physical
quantities are described by so-called local functions (LF) which are defined
on the jet space. The LF depend on $q^{a\left[  l\right]  }\;$up to some
finite orders $l\leq N_{a}\geq0$.\ The following notation is often
used\footnote{The functions $F$ may depend on time explicitly, however, we do
not include $t$ in the arguments of the functions.}:%
\begin{equation}
F\left(  q^{a\left[  0\right]  },q^{a\left[  1\right]  },q^{a\left[  2\right]
},...\right)  =F\left(  q^{[]}\right)  \label{1.2}%
\end{equation}
for the LF. In what follows, we also deal with so-called local operators (LO).
LO $\hat{U}_{Aa}$ are matrix operators which act on columns of LF $f^{a}$
producing columns $F^{A}$ of LF, $F^{A}=\hat{U}_{Aa}f^{a}\,$. LO have the form%
\begin{equation}
\hat{U}_{Aa}=\sum_{k=0}^{K<\infty}u_{Aa}^{k}\left(  d_{t}\right)  ^{k}\;,
\label{1.3}%
\end{equation}
where $u_{Aa}^{k}$ are LF. We call the operator
\begin{equation}
\left(  \hat{U}^{T}\right)  _{aA}=\sum_{k=0}^{K<\infty}\left(  -d_{t}\right)
^{k}u_{Aa}^{k} \label{1.4}%
\end{equation}
the transposed operator with respect to $\hat{U}_{Aa}$. The following relation
holds true for any LF $F^{A}$ and $f_{a}$:
\begin{equation}
F^{A}\hat{U}_{Aa}f^{a}=\left[  \left(  \hat{U}^{T}\right)  _{aA}F^{A}\right]
f^{a}+d_{t}Q\,\,, \label{1.5}%
\end{equation}
where $Q$ is an LF. The LO $\hat{U}_{ab}$ is symmetric ($+$) or antisymmetric
($-$) respectively if $\left(  \hat{U}^{T}\right)  _{ab}=\pm\hat{U}_{ab}\,.$
Thus, for any antisymmetric LO $\hat{U}_{ab}$ relation (\ref{1.5}) is reduced
to the following: $f^{a}\hat{U}_{ab}f^{b}=dQ/dt\,,$ where $Q$ is a LF.

Suppose the total time derivative of an LF vanishes. Then this LF is a
constant. Namely,
\begin{equation}
\frac{dF\left(  q^{\left[  l\right]  }\left(  t\right)  \right)  }{dt}%
\equiv0\Longrightarrow F\left(  q^{\left[  l\right]  }\right)  \equiv
\mathrm{const\,.} \label{lfa.12}%
\end{equation}
Indeed, let us suppose that $N_{a}$ are the orders of the coordinates $q^{a}$
in the LF, i.e. $F\left(  q^{\left[  l\right]  }\right)  =F\left(  \cdots
q^{a\left[  N_{a}\right]  }\right)  $. Then according to (\ref{lfa.12}) the
following relation holds true
\[
\frac{\partial F}{\partial q^{a\left[  N_{a}\right]  }}q^{a\left[
N_{a}+1\right]  }\equiv-\left[  \partial_{t}F+\sum_{a}\sum_{k=0}^{N_{a}%
-1}\frac{\partial F}{\partial q^{a\left[  k\right]  }}q^{a\left[  k+1\right]
}\right]  \,.
\]
The right hand side of the above relation does not depend on $q^{a\left[
N_{a}+1\right]  }.$ Thus, $\partial F/\partial q^{a\left[  N_{a}\right]
}\equiv0,$ and therefore $F\left(  q^{\left[  l\right]  }\right)  $ must not
depend on $q^{a\left[  N_{a}\right]  }\,$. In the same manner we can see that
$F\left(  q^{\left[  l\right]  }\right)  $ must not depend on $q^{\left[
N-1\right]  }$ and so on. If $F\left(  q^{\left[  l\right]  }\right)  $ does
not depend on any $q^{\left[  l\right]  }$ , then $\partial_{t}F\left(
q^{\left[  l\right]  }\right)  \equiv0$ as well, and we get $F\left(
q^{\left[  l\right]  }\right)  =\mathrm{const}.$

We recall that $F_{A}\left(  q^{[]}\right)  =0$ and $\chi_{\alpha}\left(
q^{[]}\right)  =0$ are equivalent sets of equations whenever they have the
same sets of solutions. In what follows, we denote this fact as
$F=0\Longleftrightarrow\chi=0\,.$ Via $O\left(  F\right)  $ we denote any LF
that vanishes on the equations $F_{a}\left(  q^{[]}\right)  =0$. More exactly,
we define $O\left(  F\right)  =\hat{V}^{b}F_{b}\,,$ where $\hat{V}^{b}$ is an
LO. Besides, we denote via $\hat{U}=\hat{O}\left(  F\right)  $ any LO that
vanish on the equations $F_{a}\left(  q^{[]}\right)  =0$. That means that the
LF $u$ that enter into (\ref{1.3}) vanish on these equations, $u=O\left(
F\right)  $, or equivalently $\hat{U}f=O\left(  F\right)  $ for any LF $f.$

We consider Lagrangian theories given by an action $S\left[  q\right]  ,$
\begin{equation}
S\left[  q\right]  =\int_{t_{1}}^{t_{2}}Ldt\,,\;L=L\left(  q^{[]}\right)  \,,
\label{1.6}%
\end{equation}
where a Lagrange function $L$ is defined as an LF on the jet
space\footnote{The functions $L$ may depend on time explicitly, however, we do
not include $t$ in the arguments of the functions.}. The Euler--Lagrange
equations are
\begin{equation}
\frac{\delta S}{\delta q^{a}}=\sum_{l=0}\left(  -d_{t}\right)  ^{l}%
\frac{\partial L}{\partial q^{a\left[  l\right]  }}=0\,. \label{1.7}%
\end{equation}

Any LF of the form $O\left(  \delta S/\delta q\right)  $ is called an extremal.

For any LF $F\left(  q^{[]}\right)  $ the operation%

\begin{equation}
\frac{d_{\text{{\scriptsize EL}}}F}{dq^{a}}=\sum_{l=0}^{N_{a}}\left(
-\frac{d}{dt}\right)  ^{l}\frac{\partial F}{\partial q^{a\left[  l\right]  }%
}\, \label{lfa.30}%
\end{equation}
is called the Euler--Lagrange derivative with respect to the coordinate\emph{
}$q^{a}$. One can see that the functional derivative of the action $S$
coincides with the Euler--Lagrange derivative of the Lagrange function,%
\begin{equation}
\frac{\delta S}{\delta q^{a}}=\frac{d_{\text{{\scriptsize EL}}}L}{dq^{a}}\,.
\label{lfa.31}%
\end{equation}
The Euler--Lagrange derivative has the following property:
\begin{equation}
\frac{d_{\text{{\scriptsize EL}}}}{dq^{a}}\frac{d}{dt}=0\,. \label{lfa.32}%
\end{equation}
To prove this, one may use the relation
\begin{align*}
&  \frac{\partial}{\partial q^{a\left[  k\right]  }}\frac{d}{dt}%
=\frac{\partial}{\partial q^{a\left[  k\right]  }}\left(  \partial_{t}%
+\sum_{l=0}q^{b\left[  l+1\right]  }\frac{\partial}{\partial q^{b\left[
l\right]  }}\right)  =(1-\delta_{k0})\frac{\partial}{\partial q^{a\left[
k-1\right]  }}\\
&  \,+\left(  \partial_{t}+\sum_{l=0}q^{b\left[  l+1\right]  }\frac{\partial
}{\partial q^{b\left[  l\right]  }}\right)  \frac{\partial}{\partial
q^{a\left[  k\right]  }}=\frac{d}{dt}\frac{\partial}{\partial q^{a\left[
k\right]  }}+(1-\delta_{k0})\frac{\partial}{\partial q^{a\left[  k-1\right]
}}\,.
\end{align*}
Thus, one gets
\begin{align*}
&  \frac{d_{\text{{\scriptsize EL}}}}{dq^{a}}\frac{d}{dt}=\sum_{k=0}\left(
-\frac{d}{dt}\right)  ^{k}\frac{\partial}{\partial q^{a\left[  k\right]  }%
}\frac{d}{dt}=-\sum_{k=0}\left(  -\frac{d}{dt}\right)  ^{k+1}\frac{\partial
}{\partial q^{a\left[  k\right]  }}+\sum_{k=1}\left(  -\frac{d}{dt}\right)
^{k}\frac{\partial}{\partial q^{a\left[  k-1\right]  }}\\
&  \,=\frac{d}{dt}\sum_{k=0}\left(  -\frac{d}{dt}\right)  ^{k}\frac{\partial
}{\partial q^{a\left[  k\right]  }}-\frac{d}{dt}\sum_{k=1}\left(
-\frac{d}{dt}\right)  ^{k-1}\frac{\partial}{\partial q^{a\left[  k-1\right]
}}=\frac{d}{dt}\frac{d_{\text{{\scriptsize EL}}}}{dq^{a}}-\frac{d}%
{dt}\frac{d_{\text{{\scriptsize EL}}}}{dq^{a}}=0\,.
\end{align*}

\subsection{Noether symmetries}

Consider an infinitesimal inner\footnote{Inner variations vanish together with
all their time\ derivatives at $t_{1}$ and $t_{2}$.} trajectory variation
$\delta q^{a}$ (inner variations vanish together with all their
time\ derivatives at $t_{1}$ and $t_{2}$). Namely,%
\begin{equation}
q^{a}\left(  t\right)  \rightarrow q^{\prime a}\left(  t\right)  =q^{a}\left(
t\right)  +\delta q^{a}\,. \label{sta.1}%
\end{equation}
We suppose that $\delta q^{a}=\delta q^{a}\left(  q^{[]}\right)  $ is an LF.
The corresponding first variation of the action can be written as follows:%
\begin{equation}
\delta S=\int_{t_{1}}^{t_{2}}\hat{\delta}L\,dt\,, \label{sta.2}%
\end{equation}
where the operator $\hat{\delta},$ which will be called the transformation
operator, acts on the corresponding LF as\footnote{Sometimes, we mark the
transformation operator below by the corresponding variation.}
\begin{equation}
\hat{\delta}=\sum_{k=0}\delta q^{a\left[  k\right]  }\frac{\partial}{\partial
q^{a\left[  k\right]  }}=\hat{\delta}_{\delta q}\,. \label{sta.3}%
\end{equation}

Two simple but useful relations follow from (\ref{sta.3}):
\begin{equation}
\hat{\delta}q^{a}=\delta q^{a}\,,\;\;\hat{\delta}_{c^{i}\delta_{i}q}=c^{i}%
\hat{\delta}_{\delta_{i}q}. \label{sta.4}%
\end{equation}

The variation (\ref{sta.1}) is a symmetry transformation of the action\emph{
}$S,$ or simply a\thinspace symmetry\thinspace of\thinspace the\thinspace
action$\emph{\;}S,$ whenever the corresponding first variation of the Lagrange
function is reduced to the total time derivative of a LF. Namely, $\delta q$
is a symmetry if%
\begin{equation}
\hat{\delta}L=\frac{dF}{dt}\,, \label{sta.5}%
\end{equation}
where $F$ is an LF. In this case the first variation (\ref{sta.2}) of the
action depends on the complete set of the variables $q^{[]}$ at $t=t_{1}$ and
$t=t_{2}$ only,
\[
\delta S=\int_{t_{1}}^{t_{2}}\hat{\delta}Ldt=\left.  F\right\vert _{t_{1}%
}^{t_{2}}\,.
\]

Any linear combination of symmetry transformations is a symmetry.

Indeed, let $\delta_{i}q$ be some symmetry transformations, and $\delta
q=c^{i}\delta_{i}q,$ where $c^{i}$ are some constants. Then, taking into
account (\ref{sta.4}), we obtain:
\begin{equation}
\hat{\delta}_{\delta_{i}q}L=\frac{dF_{i}\,}{dt}\Longrightarrow\hat{\delta
}_{\delta q}L=\frac{dF}{dt}\,,\;\;F=c^{i}F_{i}\,\,. \label{sta.6}%
\end{equation}

Transformation operators that correspond to symmetry transformations are
called symmetry operators.

The above-described symmetry transformations are called Noether symmetries.

Below, we list some properties of the transformation operators and of the
symmetry transformations:

a) Any first variation of the Lagrange function can be presented as%
\begin{equation}
\hat{\delta}L=\delta q^{a}\frac{d_{\text{{\scriptsize EL}}}L}{dq^{a}%
}+\frac{dP}{dt}=\delta q^{a}\frac{\delta S}{\delta q^{a}}+\frac{dP}{dt}\,,
\label{sta.7}%
\end{equation}
where $P$\ is an LF of the form
\begin{equation}
P=\sum\nolimits_{a}^{\prime}\sum_{m=1}^{N_{a}}p_{a}^{m}\delta q^{a\left[
m-1\right]  }\,,\;p_{a}^{m}=\sum_{s=l}^{N_{a}}\left(  -\frac{d}{dt}\right)
^{^{s-m}}\frac{\partial L}{\partial q^{a\left[  s\right]  }}\,. \label{sta.8}%
\end{equation}
One ought to remark that the sum (\ref{sta.8}) that presents $P$ is running
only over those $a$ for which $N_{a}>0.$ However, it can be extended over all
$a^{\prime}$s since the momenta $p_{a}^{m}$ that correspond to the degenerate
coordinates are zero. Thus, the prime over the sum above can be omitted.

b) Any transformation operator commutes with the total time derivative:
\begin{equation}
\left[  \hat{\delta}\,,\frac{d}{dt}\right]  =0\,. \label{sta.9}%
\end{equation}
The latter property is justified by the following relations:%
\begin{align*}
&  \frac{d}{dt}\hat{\delta}=\sum_{k=0}\left[  \delta q^{a\left[  k+1\right]
}\,\frac{\partial}{\partial q^{a\left[  k\right]  }}+\delta q^{a\left[
k\right]  }\frac{\partial}{\partial q^{a\left[  k\right]  }}\,\partial
_{t}\right]  +\sum_{k,l=0}q^{b\left[  l+1\right]  }\,\delta q^{a\left[
k\right]  }\,\frac{\partial^{2}}{\partial q^{a\left[  k\right]  }\partial
q^{b\left[  l\right]  }}\,,\\
&  \hat{\delta}\frac{d}{dt}=\sum_{l=0}\left[  \hat{\delta}q^{b\left[
l+1\right]  }\right]  \frac{\partial}{\partial q^{b\left[  l\right]  }}%
+\hat{\delta}\partial_{t}+\sum_{k,l=0}\delta q^{a\left[  k\right]
}q^{b\left[  l+1\right]  }\frac{\partial^{2}}{\partial q^{b\left[  l\right]
}\partial q^{a\left[  k\right]  }}=\frac{d}{dt}\hat{\delta}\,.
\end{align*}

c) The commutator of any two transformation operators is a transformation
operator as well.

Namely, let $\hat{\delta}_{_{1}}q=\delta q_{1}$, and $\hat{\delta}_{_{2}%
}q=\delta q_{2}$. Then%
\begin{equation}
\left[  \hat{\delta}_{1},\hat{\delta}_{2}\right]  =\hat{\delta}_{3}%
\,,\;\ \hat{\delta}_{3}q=\hat{\delta}_{1}\delta q_{2}-\hat{\delta}_{2}\delta
q_{1}\,. \label{sta.10}%
\end{equation}
Indeed, one can write:%
\begin{align}
&  \hat{\delta}_{1}\hat{\delta}_{2}=\sum_{l=0}\left(  \hat{\delta}_{1}\delta
q_{2}^{b\left[  l\right]  }\right)  \frac{\partial}{\partial q^{b\left[
l\right]  }}+\sum_{k,l=0}\delta q_{1}^{a\left[  k\right]  }\delta
q_{2}^{b\left[  l\right]  }\frac{\partial}{\partial q^{b\left[  l\right]  }%
}\frac{\partial}{\partial q^{a\left[  k\right]  }}\nonumber\\
&  =\sum_{l=0}\frac{d^{l}(\hat{\delta}_{\epsilon_{1}}\delta q_{2}^{b})}%
{dt^{l}}\frac{\partial}{\partial q^{b\left[  l\right]  }}+\sum_{k,l=0}\delta
q_{1}^{a\left[  k\right]  }\delta q_{2}^{b\left[  l\right]  }\frac{\partial
}{\partial q^{b\left[  l\right]  }}\frac{\partial}{\partial q^{a\left[
k\right]  }}\,,\label{sta.11}\\
&  \hat{\delta}_{2}\hat{\delta}_{1}=\sum_{k=0}\left(  \hat{\delta}%
_{\epsilon_{2}}\delta q_{1}^{a\left[  k\right]  }\right)  \frac{\partial
}{\partial q^{a\left[  k\right]  }}+\sum_{l,k=0}\delta q_{2}^{b\left[
l\right]  }\delta q_{1}^{a\left[  k\right]  }\frac{\partial}{\partial
q^{a\left[  k\right]  }}\frac{\partial}{\partial q^{b\left[  l\right]  }%
}\nonumber\\
&  =\sum_{k=0}\frac{d^{k}\left(  \hat{\delta}_{\epsilon_{2}}\delta q_{1}%
^{b}\right)  }{dt^{k}}\frac{\partial}{\partial q^{a\left[  k\right]  }}%
+\,\sum_{k,l=0}\delta q_{1}^{a\left[  k\right]  }\delta q_{2}^{b\left[
l\right]  }\frac{\partial}{\partial q^{b\left[  l\right]  }}\frac{\partial
}{\partial q^{a\left[  k\right]  }}\,. \label{sta.12}%
\end{align}
Then subtracting Eq. (\ref{sta.12}) from Eq. (\ref{sta.11}), we obtain the
relation (\ref{sta.10}).

In other words, the set of all transformation operators form a Lie algebra.

d) The commutator of the Euler--Lagrange derivative and a transformation
operator is proportional to the Euler--Lagrange derivative. Namely, if
$\hat{\delta}q=\delta q^{b},$ then
\begin{equation}
\left[  \frac{d_{\text{{\scriptsize EL}}}}{dq^{a}},\hat{\delta}\right]
=\hat{Q}_{a}^{b}\frac{d_{\text{EL}}}{dq^{b}}\,,\;\hat{Q}_{a}^{b}=\sum
_{k=0}\left(  -\frac{d}{dt}\right)  ^{k}\frac{\partial}{\partial q^{a\left[
k\right]  }}\delta q^{b}\,. \label{sta.13}%
\end{equation}

To prove this property, one may consider a sequence of equalities,
\begin{align*}
&  \int_{t_{1}}^{t_{2}}\frac{d_{\text{{\scriptsize EL}}}\left(  \hat{\delta
}F\right)  }{dq^{a}}\zeta^{a}dt=\int_{t_{1}}^{t_{2}}\,\hat{\delta}_{\zeta}%
\hat{\delta}Fdt=\int_{t_{1}}^{t_{2}}\hat{\delta}\hat{\delta}_{\zeta}%
Fdt+\int_{t_{1}}^{t_{2}}\hat{\delta}_{\hat{\delta}_{\zeta}\delta q}Fdt\\
&  =\int_{t_{1}}^{t_{2}}\zeta^{a}\sum_{k=0}\left(  -\frac{d}{dt}\right)
^{k}\hat{\delta}\frac{\partial F}{\partial q^{a\left[  k\right]  }}%
dt\,+\int_{t_{1}}^{t_{2}}\,\hat{\delta}_{\zeta}\delta q^{b}%
\frac{d_{\text{{\scriptsize EL}}}F}{dq^{b}}dt\\
&  =\int_{t_{1}}^{t_{2}}\,\zeta^{a}\left(  \hat{\delta}\delta_{a}^{b}+\hat
{Q}_{a}^{b}\right)  \frac{d_{\text{{\scriptsize EL}}}F}{dq^{b}}%
\,dt\,,\;\;\left(  \hat{\delta}_{\zeta}q^{a}=\zeta^{a}\right)  ,
\end{align*}
where $\zeta\left(  t\right)  $ is an arbitrary inner variation, and $F$ is an LF.

It is useful to keep in mind the following generalization of relation
(\ref{sta.13}):
\begin{equation}
\left[  \left(  \frac{d}{dt}\right)  ^{k}\frac{d_{\text{EL}}}{dq^{a}}%
,\hat{\delta}\right]  =\left(  \frac{d}{dt}\right)  ^{k}\hat{Q}_{a}%
^{b}\frac{d_{\text{{\scriptsize EL}}}}{dq^{b}}\,, \label{sta.14}%
\end{equation}
which follows immediately from (\ref{sta.9}) and (\ref{sta.13}).

e) The commutator of two symmetry operators is a symmetry operator\ as well.

Indeed, let $\hat{\delta}_{_{1}}q=\delta q_{1}$, and $\hat{\delta}_{_{2}%
}q=\delta q_{2}$ be symmetry transformations, i.e., $\hat{\delta}_{1}%
L=dF_{1}/dt\,,$ and $\hat{\delta}_{2}L=dF_{2}/dt\,.$ Then, taking into account
(\ref{sta.9}) and (\ref{sta.10}), we obtain
\begin{equation}
\left[  \hat{\delta}_{1},\hat{\delta}_{2}\right]  L=\hat{\delta}_{1}%
L=\frac{d}{dt}F_{3}\,,\;\;F_{3}=\hat{\delta}_{1}F_{2}-\hat{\delta}_{2}%
F_{1}\,\,. \label{sta.15}%
\end{equation}

Thus, the set of symmetry operators of the action $S$ forms a Lie subalgebra
of the Lie algebra of all transformation operators.

f) Symmetry transformations transform extremals into extremals.

The validity of this assertion follows from the relations proven below.

Suppose $\hat{\delta}$\ is a symmetry operator; then the following relation
takes place:
\begin{equation}
\hat{\delta}\frac{\delta S}{\delta q^{a}}=-\hat{Q}_{a}^{b}\frac{\delta
S}{\delta q^{b}}\,. \label{sta.16}%
\end{equation}
Indeed, by virtue of (\ref{lfa.31}), (\ref{lfa.32}), and (\ref{sta.13}), we
can write
\begin{align*}
&  \hat{\delta}\frac{\delta S}{\delta q^{a}}=\hat{\delta}%
\frac{d_{\text{{\scriptsize EL}}}L}{dq^{a}}=\frac{d_{\text{{\scriptsize EL}}%
}\left(  \hat{\delta}L\right)  }{dq^{a}}-\hat{Q}_{a}^{b}%
\frac{d_{\text{{\scriptsize EL}}}L}{dq^{b}}\\
&  \,=\frac{d_{\text{{\scriptsize EL}}}}{dq^{a}}\frac{dF}{dt}-\hat{Q}_{a}%
^{b}\frac{\delta S}{\delta q^{b}}=-\hat{Q}_{a}^{b}\frac{\delta S}{\delta
q^{b}}\,.
\end{align*}
A generalization of (\ref{sta.16}) based on the relation (\ref{sta.13})
reads:
\begin{equation}
\hat{\delta}\frac{d^{k}}{dt^{k}}\frac{\delta S}{\delta q^{a}}=-\frac{d^{k}%
}{dt^{k}}\hat{Q}_{a}^{b}\frac{\delta S}{\delta q^{b}}\,. \label{sta.17}%
\end{equation}

g) Symmetry transformations transform genuine trajectories into genuine trajectories.

Indeed, suppose that $\tilde{q}^{a}$ be a genuine trajectory, that is
\begin{equation}
\left.  \frac{\delta S}{\delta q^{a}}\right\vert _{\tilde{q}}=0\,,
\label{sta.18}%
\end{equation}
and $\delta q^{a}$ be a symmetry transformation. Then the transformed
trajectory $\tilde{q}^{\prime a}=\tilde{q}^{a}+\delta q^{a}$ is also a genuine
one. Indeed, by virtue of (\ref{sta.16}) and (\ref{sta.18}), we get:
\[
\left.  \frac{\delta S}{\delta q^{a}}\right\vert _{\tilde{q}^{\prime}%
=\tilde{q}+\delta q}=\left.  \frac{\delta S}{\delta q^{a}}\right\vert
_{\tilde{q}}+\left.  \hat{\delta}\frac{\delta S}{\delta q^{a}}\right\vert
_{\tilde{q}}=\left.  \left(  \delta_{a}^{b}-\hat{Q}_{a}^{b}\right)
\frac{\delta S}{\delta q^{b}}\right\vert _{\tilde{q}}=0\,.
\]

\subsection{Trivial symmetries}

Below, we are going to describe so-called\emph{\ }trivial
symmetries\ transformations, which exist for any action.

A symmetry transformation is called\emph{\ }a trivial symmetry
transformation\emph{ }whenever the corresponding trajectory variation has the
form
\begin{equation}
\delta q^{a}=\hat{U}^{ab}\frac{\delta S}{\delta q^{b}}\,, \label{sta.19}%
\end{equation}
where $\hat{U}$ is an antisymmetric LO, that is $\left(  \hat{U}^{T}\right)
^{ab}=-\hat{U}^{ab}$ . Thus, trivial symmetry transformations do not affect
genuine trajectories. (One can prove, see below, that any symmetry
transformation that vanishes on the equations of motion, $\delta
q^{a}=O\left(  \delta S/\delta q\right)  ,$ is trivial, namely it has the form
(\ref{sta.19})). With the help of relations (\ref{1.5}) and (\ref{sta.7}), we
can easily verify that (\ref{sta.19}) is actually a symmetry transformation.
Indeed,
\[
\hat{\delta}L=\frac{d_{\text{{\scriptsize EL}}}L}{dq^{b}}\hat{U}^{ab}%
\frac{d_{\text{{\scriptsize EL}}}L}{dq^{b}}+\frac{dP}{dt}=\frac{dF}%
{dt}\,+\frac{dP}{dt}=\frac{d\left(  F+P\right)  }{dt}\,,
\]
where $F$ and $P$ are some LF.

Since trivial symmetry transformations are proportional to the EM, they do not
change genuine trajectories, as was already mentioned above.

The commutator of a symmetry operator and a trivial-symmetry operator is a
trivial-symmetry operator. Namely, if
\[
\hat{\delta}_{1}L=dF_{1}/dt\,,\;\hat{\delta}_{2}L=dF_{2}/dt,\;\hat{\delta}%
_{2}q^{a}=\delta_{2}q^{a}=\hat{V}^{ab}\delta S/\delta q^{b}\,,
\]
then
\begin{equation}
\left[  \hat{\delta}_{1},\hat{\delta}_{2}\right]  L=\hat{\delta}_{3}%
L\,,\;\hat{\delta}_{3}q^{a}=\delta_{3}q^{a}=\hat{U}^{ab}\frac{\delta S}{\delta
q^{b}}\,, \label{sta.20}%
\end{equation}
where $\hat{V}^{ab}\;$and $\hat{U}^{ab}$ are some antisymmetric LO.

To verify (\ref{sta.20}), we remark that, according to (\ref{sta.10}),
$\hat{\delta}_{3}$ is a symmetry operator, with $\delta_{3}q=\hat{\delta}%
_{1}\delta_{2}q-\hat{\delta}_{2}\delta_{1}q,$ \ where $\delta_{1}q=\hat
{\delta}_{1}q^{a}.$ The term $\hat{\delta}_{1}\delta_{2}q$ can be calculated
with the help of (\ref{sta.3}),
\[
\hat{\delta}_{1}\delta_{2}q^{a}=\sum_{k=0}\frac{\partial\left(  \delta
_{2}q^{a}\right)  }{\partial q^{c\left[  k\right]  }}\left[  \frac{d^{k}%
}{dt^{k}}\left(  \hat{V}^{cb}\frac{\delta S}{\delta q^{b}}\right)  \right]
\,,
\]
and the term $\hat{\delta}_{2}\delta_{1}q$ can be calculated with the help of
(\ref{sta.16}),
\[
\hat{\delta}_{2}\delta_{1}q=\left(  \hat{\delta}_{2}\hat{V}^{ab}\right)
\frac{\delta S}{\delta q^{b}}+\hat{V}^{ab}\hat{\delta}_{2}\frac{\delta
S}{\delta q^{b}}\,=\left(  \hat{\delta}_{2}\hat{V}^{ab}\right)  \frac{\delta
S}{\delta q^{b}}-\hat{V}^{ab}\hat{Q}_{b}^{c}\frac{\delta S}{\delta q^{c}}\,.
\]
Thus, we obtain: $\hat{\delta}_{3}q^{a}=\delta_{3}q^{a}=\hat{U}^{ab}\delta
S/\delta q^{b},$ where $\hat{U}^{ab}$ is an antisymmetric LO of the form
\[
\hat{U}^{ab}=\sum_{k=0}\left[  \frac{\partial\left(  \delta_{2}q^{a}\right)
}{\partial q^{c\left[  k\right]  }}\left(  \frac{d}{dt}\right)  ^{k}\hat
{V}^{cb}+\hat{V}^{ac}\left(  -\frac{d}{dt}\right)  ^{k}\frac{\partial\left(
\delta_{2}q^{b}\right)  }{\partial q^{c\left[  k\right]  }}\right]
-\hat{\delta}_{2}\hat{V}^{ab}.
\]

We call two symmetry transformations $\delta_{1}q$ and $\delta_{2}q$
equivalent ($\delta_{1}q\sim\delta_{2}q$)\ whenever they differ by a trivial
symmetry transformation:
\begin{equation}
\delta_{1}q\sim\delta_{2}q\Longleftrightarrow\delta_{1}q^{a}-\delta_{2}%
q^{a}=\hat{U}^{ab}\frac{\delta S}{\delta q^{b}}\,. \label{sta.21}%
\end{equation}
Here $\left(  \hat{U}^{T}\right)  ^{ab}=-\hat{U}^{ab}\,.$

Let $\mathbf{G}\left(  S\right)  $ be the Lie algebra of all symmetries of the
action $S$. The trivial symmetries form the ideal $\mathbf{G}_{\mathrm{tr}%
}\left(  S\right)  $\ in the Lie algebra $\mathbf{G}\left(  S\right)  $. Then
the classes of equivalent symmetries form a Lie algebra $\mathbf{G}%
_{\mathrm{Ph}}\left(  S\right)  $ isomorphic to the quotient algebra:
\[
\mathbf{G}_{\mathrm{Ph}}\left(  S\right)  =\mathbf{G}\left(  S\right)
/\mathbf{G}_{\mathrm{tr}}\left(  S\right)  \,.
\]

\section{Dynamically equivalent actions}

Very often we encounter an action%
\begin{equation}
S_{\mathrm{E}}[q,y]=\int L_{\mathrm{E}}\left(  q^{[]},y^{[]}\right)  dt\,,
\label{std.1}%
\end{equation}
which contains two groups of coordinates $q^{[]}$ and $y^{[]}$ such that the
Euler--Lagrange allow one to express all $y$ via $q^{[]}.$ It is convenient to
call $S_{\mathrm{E}}[q,y]$ the extended action. One can try to eliminate the
variables $y$ from the extended action to get some reduced action, which
depends now only on $q$, and ask the question: What is the relation between
the extended and the reduced actions? There exist a case when this question
has a definite answer \cite{2,5}. Namely, let us suppose that the
Euler--Lagrange $\delta S_{\mathrm{E}}\left[  q,y\right]  /\delta y=0$ allow
one to express uniquely the variables $y$ as LF of the variables $q$,%
\begin{equation}
\frac{\delta S_{\mathrm{E}}\left[  q,y\right]  }{\delta y}%
=0\Longleftrightarrow y=\bar{y}\left(  q^{[]}\right)  \,. \label{lfb.7}%
\end{equation}
Then we define the reduced action $S\left[  q\right]  $%
\begin{equation}
S\left[  q\right]  =S_{\mathrm{E}}[q,\bar{y}]=\int L_{\mathrm{E}}\left(
q^{[]},\bar{y}^{[]}\right)  dt=\int L\left(  q^{[]}\right)  dt\,.
\label{std.3}%
\end{equation}
Let us compare the Euler--Lagrange that correspond to both actions. First
consider the variation of the reduced action $\delta S\,\ $under arbitrary
inner variations $\delta q$,%
\begin{equation}
\delta S[q]=\int\left(  \left.  \frac{\delta S_{\mathrm{E}}\left[  q,y\right]
}{\delta q^{i}}\right|  _{y=\bar{y}}\delta q^{i}+\left.  \frac{\delta
S_{\mathrm{E}}\left[  q,y\right]  }{\delta y^{\alpha}}\right|  _{y=\bar{y}%
}\delta\bar{y}^{\alpha}\right)  dt=\int\frac{\delta S\left[  q\right]
}{\delta q^{i}}\delta q^{i}dt\,. \label{lfb.8}%
\end{equation}
In virtue of (\ref{lfb.7}), the Euler--Lagrange of the reduced action read%
\begin{equation}
\frac{\delta S\left[  q\right]  }{\delta q}=\left.  \frac{\delta
S_{\mathrm{E}}\left[  q,y\right]  }{\delta q}\right|  _{y=\bar{y}}=0\,.
\label{lfb.9}%
\end{equation}
On the other hand, the Euler--Lagrange of the extended action $S_{\mathrm{E}%
}\left[  q,y\right]  $ are%
\[
\frac{\delta S_{\mathrm{E}}\left[  q,y\right]  }{\delta q}=0\,,\;\frac{\delta
S_{\mathrm{E}}\left[  q,y\right]  }{\delta y}=0\Longleftrightarrow y=\bar
{y}\left(  q^{[]}\right)  \;.
\]
They are reduced to (\ref{lfb.9}) in the $q$-sector. We can see that the
extended action and the reduced action lead to the same Euler--Lagrange for
$q$. This is why the variables $y$ are called the\ auxiliary variables. The
auxiliary variables $y$ can be eliminated from the action with the help of the
Euler--Lagrange. Further, we call the actions $S_{\mathrm{E}}\left[
q,y\right]  $ and $S[q]$ the dynamically equivalent actions.

One ought to stress that the above equivalence is a consequence of the
assumption that the variables $y$ are expressed via $q$ by means of the
equations $\delta S/\delta y=0$ only. If, for this purpose, some of the
equations $\delta S/\delta q=0$ are used as well, then the above equivalence
can be absent. Of course, the solutions of the Euler--Lagrange for the reduced
action, together with the definition $y=\bar{y},\;$contain all solutions of
the Euler--Lagrange for the extended action (as it is easily seen from Eq.
(\ref{lfb.8})). However, the reduced action can imply additional solutions.

Actions containing auxiliary variables and the corresponding reduced actions
have similar properties, in particular, there exists a direct relation between
their symmetry transformations.

As was mentioned above, we are going to relate the symmetry properties of the
extended and reduced actions. To this end, it is convenient to make an
invertible coordinate replacement, $\left(  q^{a},y^{\alpha}\right)
\rightarrow\tilde{q}^{A}=\left(  q^{a},z^{\alpha}\right)  $, $y=z+\bar
{y}\left(  q^{\left[  l\right]  }\right)  ,$ in the extended action. In fact,
we are going to consider a modified extended action $\tilde{S}[\tilde{q}],$
which is obtained from the extended action $S_{\mathrm{E}}[q,y]$ as follows:%
\begin{equation}
\tilde{S}[\tilde{q}]=\int\tilde{L}\left(  \tilde{q}^{[]}\right)
dt=S_{\mathrm{E}}[q,z+\bar{y}]=\int L_{\mathrm{E}}\left(  q^{[]},z^{[]}%
+\bar{y}^{[]}\right)  dt\,. \label{std.4}%
\end{equation}
The extended action $S_{\mathrm{E}}[q,y]$ and the modified extended action
$\tilde{S}[\tilde{q}]$ are completely equivalent. They lead to completely
equivalent Euler--Lagrange. Thus, it is sufficient to study the relation
between the symmetry properties of the modified extended action $\tilde
{S}[\tilde{q}]$ and the reduced action $S\left[  q\right]  .$

Note that
\begin{equation}
S\left[  q\right]  =\left.  \tilde{S}[\tilde{q}]\right|  _{z=0}\,,\;L\left(
q^{[]}\right)  =\left.  \tilde{L}\left(  \tilde{q}^{[]}\right)  \right|
_{z=0}\,. \label{std.5}%
\end{equation}
Besides, the action (\ref{std.4}) can be presented in the form
\begin{align}
&  \tilde{S}[\tilde{q}]=S\left[  q\right]  +\Delta S\left[  \tilde{q}\right]
\,,\;\;\Delta S\left[  \tilde{q}\right]  =\int\Delta Ldt\,,\nonumber\\
&  \Delta L=\tilde{L}\left(  \tilde{q}^{[]}\right)  -L\left(  q^{[]}\right)
=L_{\mathrm{E}}\left(  q^{[]},z^{[]}+\bar{y}^{[]}\right)  -L_{\mathrm{E}%
}\left(  q^{[]},\bar{y}^{[]}\right)  \,. \label{std.6}%
\end{align}
The variables $z$ are auxiliary ones for the action $\tilde{S}[\tilde{q}],$
and, in particular, $z=0$ on the Euler--Lagrange. Indeed,
\begin{equation}
\frac{\delta\tilde{S}[\tilde{q}]}{\delta z}=0\Longleftrightarrow\frac{\delta
S_{\mathrm{E}}[q,y]}{\delta y}=0\Longrightarrow y=\bar{y}\left(
q^{[]}\right)  \Longrightarrow z=0\,. \label{std.7}%
\end{equation}
The latter implies:
\begin{equation}
\frac{\delta\tilde{S}}{\delta z^{\alpha}}=\frac{\delta\Delta S}{\delta
z^{\alpha}}=\hat{U}_{\alpha\beta}z^{\beta}=0\,.\, \label{std.8}%
\end{equation}
Since equation (\ref{std.7}) has the unique solution $z=0$, one can easily
verify that $\hat{U}$ is an invertible LO. The equation (\ref{std.8}) implies
\begin{equation}
\Delta L=z^{\alpha}\hat{K}_{\alpha\beta}z^{\beta}+\frac{d}{dt}F\,,
\label{std.9}%
\end{equation}
where $\hat{K}$ is a symmetric LO, and $F$ is an LF. Besides, one can write
\begin{equation}
z^{\alpha}=\left(  \hat{U}^{-1}\right)  ^{\alpha\beta}\frac{\delta\Delta
S}{\delta z^{\beta}}=\left(  \hat{U}^{-1}\right)  ^{\alpha\beta}%
\frac{\delta\tilde{S}}{\delta z^{\beta}}\,. \label{std.9a}%
\end{equation}
On the other hand, due to the property (\ref{lfa.32}), one can write
\[
\frac{\delta\Delta S}{\delta q^{a}}=\frac{d_{\text{{\scriptsize EL}}}\Delta
L}{dq^{a}}=\frac{d_{\text{{\scriptsize EL}}}}{dq^{a}}\left[  z^{\alpha}\hat
{K}_{\alpha\beta}z^{\beta}\right]  \,.
\]
Then, taking into account (\ref{std.9}, \ref{std.9a}), and the definition of
the Euler--Lagrange derivative, we get the following useful relation:
\begin{equation}
\frac{\delta\Delta S}{\delta q^{a}}=\hat{\Lambda}_{a}^{\alpha}\frac{\delta
\Delta S}{\delta z^{\alpha}}\,,\;\;\hat{\Lambda}_{a}^{\alpha}=\sum
_{l=0}\left(  -\frac{d}{dt}\right)  ^{l}z^{\nu}\frac{\partial\hat{K}_{\nu
\beta}}{\partial q^{a\left[  l\right]  }}\left(  \hat{U}^{-1}\right)
^{\beta\alpha}\,, \label{std.10}%
\end{equation}
where $\hat{\Lambda}_{a}^{\alpha}$ is an LO.

\section{Symmetries of the extended and the reduced actions}

There exists a one-to-one correspondence (isomorphism) between the symmetry
classes of the extended action $\tilde{S}[\tilde{q}]$ and the reduced action
$S\left[  q\right]  .$ Below, we prove a set of assertions, which justify, in
fact, this correspondence.

i) If the transformation%
\begin{equation}
\delta\tilde{q}^{A}=\left(
\begin{array}
[c]{c}%
\delta^{\prime}q^{a}\\
\delta z^{\alpha}%
\end{array}
\right)  , \label{std.11}%
\end{equation}
is a symmetry of the extended action $\tilde{S},$ then the transformation%
\begin{equation}
\delta q^{a}=\left.  \delta^{\prime}q\right|  _{z=0}\, \label{std.12}%
\end{equation}
is a symmetry of the reduced action $S.$

Indeed, let (\ref{std.11}) be a symmetry of the action $\tilde{S}$. Then
\begin{equation}
\hat{\delta}_{\delta\tilde{q}}\tilde{L}=\frac{d}{dt}\tilde{F}\,,
\label{std.13}%
\end{equation}
where $\tilde{F}$ is an LF. Considering (\ref{std.13}) at $z=\delta z=0,$ we
get
\[
\hat{\delta}_{\delta q}L=\frac{d}{dt}F\,,\;\;\delta q^{a}=\left.
\delta^{\prime}q\right\vert _{z=0}\,,\;\;F=\left.  \tilde{F}\right\vert
_{z=0}\;,
\]
where $L$ is given by (\ref{std.5}). Thus, any symmetry of the action
$\tilde{S}$ implies a symmetry of the action $S$. The symmetry $\delta q$
obtained in such a way can be called the symmetry reduction of the extended action.

ii) If the transformation $\delta q$ is a symmetry of the reduced action $S$,
then the transformation
\begin{equation}
\delta\tilde{q}^{A}=\left(
\begin{array}
[c]{c}%
\delta q^{a}\\
\delta z^{\alpha}%
\end{array}
\right)  ,\;\delta z^{\alpha}=-\left(  \hat{\Lambda}^{T}\right)  _{a}^{\alpha
}\delta q^{a}\,, \label{std.15}%
\end{equation}
where the LO $\hat{\Lambda}$ defined by Eq. (\ref{std.10}) is a symmetry of
the extended action $\tilde{S}.$

To prove this assertion, let us consider the first variation $\hat{\delta
}_{\delta\tilde{q}}\tilde{L}$ of the Lagrange function $\tilde{L}$ . Since
$\delta q$ is a symmetry of the reduced action $S$, the relation $\hat{\delta
}_{\delta q}L=dF/dt$ , where $F$ is an LF, holds true. Thus, with the help of
the property (\ref{sta.4}), one may write the variation $\hat{\delta}%
_{\delta\tilde{q}}\tilde{L}$ in the form
\begin{equation}
\hat{\delta}_{\delta\tilde{q}}\tilde{L}=\left(  \hat{\delta}_{\delta q}%
+\hat{\delta}_{\delta z}\right)  \tilde{L}=\frac{d}{dt}F+\left(  \hat{\delta
}_{\delta q}+\hat{\delta}_{\delta z}\right)  \Delta L\,. \label{std.16}%
\end{equation}
Now, we present the variations $\hat{\delta}_{\delta q}\Delta L$ and
$\hat{\delta}_{\delta z}\Delta L$ with the help of relation (\ref{sta.7}).
Besides, taking into account the expression (\ref{std.15}) for the variation
$\delta z,$ we get
\begin{equation}
\hat{\delta}_{\delta\tilde{q}}\tilde{L}=\frac{d}{dt}\left(  F+P_{q}%
+P_{z}\right)  +\delta q^{a}\frac{\delta\Delta S}{\delta q^{a}}-\left[
\left(  \hat{\Lambda}^{T}\right)  _{a}^{\alpha}\delta q^{a}\right]
\frac{\delta\Delta S}{\delta z^{\alpha}}\,, \label{std.17}%
\end{equation}
where $P_{q}$ and $P_{z}$ are some LF. Using (\ref{std.10}) and (\ref{1.5}),
we may write%
\begin{equation}
\delta q^{a}\frac{\delta\Delta S}{\delta q^{a}}=\delta q^{a}\hat{\Lambda}%
_{a}^{\alpha}\frac{\delta\Delta S}{\delta z^{\alpha}}=\left[  \left(
\hat{\Lambda}^{T}\right)  _{a}^{\alpha}\delta q^{a}\right]  \frac{\delta\Delta
S}{\delta z^{\alpha}}+\frac{dG}{dt}\,, \label{std.18}%
\end{equation}
where $G$ is an LF. Thus, the variation $\hat{\delta}_{\delta\tilde{q}}%
\tilde{L}$ is reduced to the total derivative of an LF,
\[
\hat{\delta}_{\delta\tilde{q}}\tilde{L}=\frac{d}{dt}\left(  F+P_{q}%
+P_{z}+G\right)  \,.
\]
Thus, $\delta\tilde{q}$ is a symmetry of the extended action $\tilde{S}$.

iii) Any symmetry of the form
\begin{equation}
\delta\tilde{q}=\left(
\begin{array}
[c]{c}%
0\\
\delta z
\end{array}
\right)  \label{std.25a}%
\end{equation}
of the extended action $\tilde{S}$ is trivial.

Since $\delta\tilde{q}$ is a symmetry of the action $\tilde{S},$ one can
write
\begin{equation}
\hat{\delta}_{\delta\tilde{q}}\tilde{L}=\hat{\delta}_{\delta z}\tilde
{L}=\frac{dF}{dt}\,, \label{std.25}%
\end{equation}
where $F$ is an LF. Taking into account (\ref{sta.7}), we may rewrite Eq.
(\ref{std.25}) as
\begin{equation}
\delta z^{\alpha}\frac{\delta\tilde{S}}{\delta z^{\alpha}}=\frac{dF^{\prime}%
}{dt}\,, \label{std.26}%
\end{equation}
where $F^{\prime}$ is an LF. The left-hand side of equation (\ref{std.26}) can
be transformed, with the help of (\ref{std.8}) and (\ref{1.5}), to the form
\[
\delta z^{\alpha}\frac{\delta\tilde{S}}{\delta z^{\alpha}}=\delta z^{\alpha
}\hat{U}_{\alpha\beta}z^{\beta}=\left[  \left(  \hat{U}^{T}\right)
_{\beta\alpha}\sigma^{\alpha}\right]  z^{\beta}+\frac{dF^{\prime\prime}}{dt},
\]
where $F^{\prime\prime}$ is an LF. Thus, the equation (\ref{std.26}) may be
reduced to%
\begin{equation}
z^{\beta}f_{\beta}=\frac{d\Phi}{dt}\,,\;f_{\beta}=\left(  \hat{U}^{T}\right)
_{\beta\alpha}\sigma^{\alpha}\,, \label{std.27}%
\end{equation}
where $f\left(  Q^{[]}\right)  $ and $\Phi\left(  Q^{[]}\right)  $ are some
LF. Let us present the LF $\Phi$ as%
\begin{align}
&  \Phi\left(  Q^{[]}\right)  =\Phi_{0}\left(  q^{[]}\right)  +\Phi_{1}\left(
Q^{[]}\right)  \,,\nonumber\\
&  \Phi_{0}=\left.  \Phi\right|  _{z=0}\,,\;\left.  \Phi_{1}\right|
_{z=0}=\sum_{k=0}^{N}\Phi_{\alpha\left(  k\right)  }\left(  Q^{[]}\right)
z^{\alpha\left[  k\right]  }\,,\;N<\infty\,. \label{std.28}%
\end{align}
It follows from equation (\ref{std.27}) that $d\Phi_{0}/dt\equiv0.$ According
to (\ref{lfa.12}), the latter implies $\Phi_{0}\equiv\mathrm{const}\,.$ From
(\ref{std.27}), we get the equation%
\begin{equation}
\sum_{k=0}^{N+1}\varphi_{\alpha\left(  k\right)  }z^{\alpha\left[  k\right]
}=0\,, \label{std.30}%
\end{equation}
where
\begin{align}
&  \varphi_{\alpha\left(  0\right)  }=f_{\alpha}-\dot{\Phi}_{\alpha\left(
0\right)  \,},\;\varphi_{\alpha\left(  N+1\right)  }=-\Phi_{\alpha\left(
N\right)  }\,,\nonumber\\
&  \varphi_{\alpha\left(  k\right)  }=-\left[  \Phi_{\alpha\left(  k-1\right)
}+\dot{\Phi}_{\alpha\left(  k\right)  }\right]  ,\;k=1,...,N\,. \label{std.31}%
\end{align}
The general solution of Eq. (\ref{std.30}) is%

\begin{equation}
\varphi_{\alpha\left(  k\right)  }=\sum_{l=0}^{N+1}m_{\alpha\left(  k\right)
|\beta\left(  s\right)  l}\,z^{\beta\left[  s\right]  },\;\;m_{\alpha\left(
k\right)  |\beta\left(  s\right)  l}=-m_{\beta\left(  s\right)  |l\alpha
\left(  k\right)  }\,, \label{std.32}%
\end{equation}
where $m_{\alpha\left(  k\right)  |\beta\left(  s\right)  l}\left(
Q^{[]}\right)  $ are some LF. Then the LF $\Phi_{\alpha\left(  k\right)  }$
and $f_{\alpha}$\ can be found from Eq. (\ref{std.31}):
\begin{align}
&  \Phi_{\alpha\left(  k\right)  }=-\sum_{m=0}^{N-k}\sum_{l=0}^{N+1}\left(
-\frac{d}{dt}\right)  ^{m}\left[  m_{\alpha\left(  k+m+1\right)  |\beta\left(
l\right)  }z^{\beta\left[  l\right]  }\right]  \,\,,\nonumber\\
&  f_{\alpha}=\sum_{m,l=0}^{N+1}\left(  -\frac{d}{dt}\right)  ^{m}\left[
m_{\alpha\left(  m\right)  |\beta\left(  l\right)  }z^{\beta\left[  l\right]
}\right]  \equiv\hat{m}_{\alpha\beta}z^{\beta}\,, \label{std.29}%
\end{align}
where $\hat{m}_{\alpha\beta}$ is an antisymmetric LO. Thus, we get from
(\ref{std.27})
\begin{equation}
\delta z^{\alpha}=\hat{M}^{\alpha\beta}\frac{\delta\tilde{S}}{\delta z^{\beta
}},\;\hat{M}^{\alpha\beta}=\left[  \left(  \hat{U}^{T}\right)  ^{-1}\right]
^{\alpha\gamma}\hat{m}_{\gamma\delta}\left(  \hat{U}^{-1}\right)
^{\delta\beta}\,, \label{std.33}%
\end{equation}
where $\hat{M}^{\alpha\beta}$ is an antisymmetric LO. Therefore, the symmetry
(\ref{std.25a}) is trivial.

iv) Suppose both transformations $\delta\tilde{q}_{1}$ and $\delta\tilde
{q}_{2}$ to be symmetries of the extended action $\tilde{S}$\ such that their
reductions coincide, that is
\begin{equation}
\left.  \delta^{\prime}q_{1}\right\vert _{z=0}=\left.  \delta^{\prime}%
q_{2}\right\vert _{z=0}=\delta q\,. \label{std.19}%
\end{equation}
Then these symmetries are equivalent,
\begin{equation}
\delta\tilde{q}_{1}\sim\delta\tilde{q}_{2}\,, \label{std.20}%
\end{equation}
which means that $\delta\tilde{q}_{1}$ and $\delta\tilde{q}_{2}$\ differ by a
trivial symmetry.

Thus, we have to proved that the transformation
\[
\Delta\tilde{q}=\delta\tilde{q}_{1}\mathbf{-}\delta\tilde{q}_{2}=\left(
\begin{array}
[c]{l}%
\Delta q^{\prime}=\delta^{\prime}q_{1}-\delta^{\prime}q_{2}\\
\Delta z=\delta z_{1}-\delta z_{2}%
\end{array}
\right)  ,\;\left.  \Delta q^{\prime}\right\vert _{z=0}=0\,,
\]
is a trivial symmetry of the extended action $\tilde{S}\,.$ In virtue of Eq.
(\ref{std.19}), the LF $\Delta q^{\prime}$ may be presented as
\begin{equation}
\Delta q^{\prime a}=\hat{m}_{\alpha}^{a}z^{\alpha}\,, \label{std.21}%
\end{equation}
where $\hat{m}$ is an LO. With the help of (\ref{std.9a}), we get for $\Delta
q^{\prime}$ the following expression:
\begin{equation}
\Delta q^{\prime a}=\hat{M}^{a\beta}\frac{\delta\tilde{S}}{\delta z^{\beta}%
}\,, \label{std.22}%
\end{equation}
where $\hat{M}=\hat{m}\hat{U}^{-1}$ is an LO.

Let us present the transformation $\Delta\tilde{q}$ in the form $\Delta
\tilde{q}=\Delta_{1}\tilde{q}+\Delta_{2}\tilde{q},$ where
\begin{equation}
\Delta_{1}\tilde{q}=\hat{M}^{AB}\frac{\delta\tilde{S}}{\delta\tilde{q}^{B}%
}\,,\;\hat{M}^{AB}=\left(
\begin{array}
[c]{cc}%
0 & \hat{M}^{a\beta}\\
-\left(  \hat{M}^{T}\right)  ^{\alpha b} & 0
\end{array}
\right)  , \label{std.23}%
\end{equation}
and
\begin{equation}
\Delta_{2}\tilde{q}=\left(
\begin{array}
[c]{c}%
0\\
\Delta\sigma^{\prime\prime}%
\end{array}
\right)  . \label{std.24}%
\end{equation}
The transformations $\Delta_{1}\tilde{q}$ is a trivial symmetry since the LO
$\hat{M}^{AB}$ is antisymmetric, that is $\left(  \hat{M}^{T}\right)
^{AB}=-\hat{M}^{AB}$. Thus, $\Delta_{2}\tilde{q}$ is a symmetry of the
extended action $\tilde{S}.$ Besides, the latter symmetry has a special form
(\ref{std.24}). It was proven in item c) that any symmetry of such a form is
trivial. Therefore, the symmetry $\Delta\tilde{q}$ is trivial as well.

v) Let a transformation $\delta\tilde{q}$ be a trivial symmetry of the
extended action $\tilde{S}.$ Then its reduction $\delta q$ is a trivial
symmetry of the reduced action $S.$

According to this assumption, we may write
\begin{equation}
\delta\tilde{q}^{A}=\left(
\begin{array}
[c]{l}%
\delta^{\prime}q^{a}=\hat{M}^{ab}\frac{\delta\tilde{S}}{\delta q^{b}}+\hat
{M}^{a\beta}\frac{\delta\tilde{S}}{\delta z^{\beta}}\\
\delta z^{\alpha}=-\left(  \hat{M}^{T}\right)  ^{b\alpha}\frac{\delta\tilde
{S}}{\delta q^{b}}+\hat{M}^{\alpha\beta}\frac{\delta\tilde{S}}{\delta
z^{\beta}}%
\end{array}
\right)  , \label{std.34}%
\end{equation}
where the local operators $\hat{M}^{ab}$ and $\hat{M}^{\alpha\beta}$ are
antisymmetric.\ Then the reduction $\delta q=\left.  \delta^{\prime
}q\right\vert _{z=0}$ of the transformation (\ref{std.34}) reads%
\begin{equation}
\delta q^{a}=\hat{m}^{ab}\frac{\delta S}{\delta q^{b}},\;\hat{m}^{ab}=\left.
\hat{M}^{ab}\right\vert _{z=0}\;. \label{std.35}%
\end{equation}
The LO $\hat{m}^{ab}$ is antisymmetric. Thus, (\ref{std.35}) is a trivial
symmetry of the reduced action $S$.

vi) Let a symmetry $\delta q$ of a reduced action $S$ be trivial. Then any
extension of this symmetry to the symmetry $\delta\tilde{q}$ of the extended
action $\tilde{S}$ is trivial as well.

Since $\delta q$ is a trivial symmetry, one can write%
\[
\delta q^{a}=\hat{m}^{ab}\frac{\delta S}{\delta q^{b}}\,,
\]
where $\hat{m}^{ab}$ is an antisymmetric LO. Consider the following extension
of the symmetry $\delta q^{a}$:%
\begin{equation}
\delta\tilde{q}_{1}=\left(
\begin{array}
[c]{c}%
\delta^{\prime}q\\
0
\end{array}
\right)  ,\;\delta^{\prime}q^{a}=\hat{m}^{ab}\frac{\delta\tilde{S}}{\delta
q^{b}}\,, \label{std.37}%
\end{equation}
which is a trivial symmetry of the extended action $\tilde{S}.$ Any other
extension of $\delta q$ differs from $\delta\tilde{q}_{1}$ by a trivial
symmetry, according to item (iv). Therefore, any extension of the trivial
symmetry is a trivial symmetry as well.

Concluding, we can see that there exists an isomorphism between classes of
equivalent symmetries of dynamically equivalent actions. Since the Lagrangian
and Hamiltonian actions are dynamically equivalent, one can study the symmetry
structure of any singular theory considering the first-order Hamiltonian action.

\begin{acknowledgement}
Gitman is grateful to the foundations FAPESP, CNPq for permanent support and
to the Lebedev Physics Institute (Moscow) for hospitality; Tyutin thanks RFBR
05-02-17217 and LSS-1578-2003.2 for partial support.
\end{acknowledgement}

\end{document}